\documentclass[runningheads]{llncs}
\pdfoutput=1
\usepackage[T1]{fontenc}
\usepackage{amsmath}
\usepackage{graphicx}
\usepackage{booktabs}
\usepackage{amssymb}
\usepackage{multirow}
\usepackage[hyphens]{url}
\newcommand\etal{\emph{et al.}}

\begin{document}
\title{Plaintext-Free Deep Learning for Privacy-Preserving Medical Image Analysis via Frequency Information Embedding}
%\title{Not Using Plaintext Medical Images in Training and Inference Stages of DL Models}
%\title{Redefining Medial Image Data Use in Deep Learning: The Case for Non-Plaintext Medical Images}
%\title{Transforming Medical Image Analysis in Deep Learning ：Utilizing Non-Plaintext Images for Privacy Preservation
%Anonymous Medical Image Analysis

% \author{Mengyu Sun\inst{1} \and Ziyuan Yang\inst{2} \and Yi Zhang\inst{2}}
% \institute{ School of Cyber Science and Engineering, Chengdu 610065, China \and
 % College of Computer Science, Sichuan University, Sichuan University, Chengdu 610065, China\\
 % \email{x@x.com}}

\author{Mengyu Sun\inst{1}
Ziyuan Yang\inst{2} \and
Maosong Ran\inst{2} \and
Zhiwen Wang\inst{2} \and
Hui Yu\inst{2} \and Yi~Zhang\inst{1}}

\authorrunning{M. Sun et al.}

\institute{School of Cyber Science and Engineering, Sichuan University \and
College of Computer Science, Sichuan University \\
Corresponding Author: Yi Zhang (email: \email{yzhang@scu.edu.cn})
}

 % \authorrunning{Anonymous}
 % \institute{Anonymous Institution}
%
\maketitle              % typeset the header of the contribution
%
% \vspace{-10pt}
\begin{abstract}
In the fast-evolving field of medical image analysis, Deep Learning (DL)-based methods have achieved tremendous success. However, these methods require plaintext data for training and inference stages, raising privacy concerns, especially in the sensitive area of medical data. 
To tackle these concerns, this paper proposes a novel framework that uses surrogate images for analysis, eliminating the need for plaintext images. This approach is called Frequency-domain Exchange Style Fusion (FESF). The framework includes two main components: Image Hidden Module (IHM) and Image Quality Enhancement Module~(IQEM). The~IHM performs in the frequency domain, blending the features of plaintext medical images into host medical images, and then combines this with IQEM to improve  and create surrogate images effectively. During the diagnostic model training process, only surrogate images are used, enabling anonymous analysis without any plaintext data during both training and inference stages. Extensive evaluations demonstrate that our framework effectively preserves the privacy of medical images and maintains diagnostic accuracy of DL models at a relatively high level, proving its effectiveness across various datasets and DL-based models.

\keywords{Privacy-preserving  \and Medical Image analysis \and Deep Learning \and Frequency-Domain Exchange \and Image concealment.}
\end{abstract}

\section{Introduction}
Rapidly developing deep learning (DL) methodologies have demonstrated great potential in the medical field, offering high accuracy and efficiency in diagnosis, treatment planning, and patient care~\cite{zhou2021review,padhi2023transforming}.
However, as DL is a data-driven approach~\cite{he2019model,yang2023hypernetwork}, the increasing demand for large volumes of plaintext medical images for training DL models raises significant privacy concerns in its applications within the medical field.
Medical images hold extensive privacy information, including personal identification, medical history, diagnosis, biometric data, and more~\cite{kaissis2020secure}. 
Therefore, to facilitate the public availability of large-scale medical image databases for the purposes of advancing medical research and improving patient care, the development and implementation of robust privacy-preserving methods for medical imaging have become urgently necessary. 

Traditional DL-based methodologies necessitate the server's aggregation of data from multiple clients, raising privacy concerns during client-server data transmission. These methods also rely on plaintext medical images for training and inference~\cite{belal2022comprehensive,aminizadeh2023applications}. With this scenario, the application of DL methodologies to analyze anonymized medical images has emerged as a promising approach to overcome data-sharing limitations and mitigating the risk of data leakage. Traditional anonymization techniques for medical images often involve removing the patient's name or blurring identifiable information. Some DL-based methods, such as Kai~\etal~\cite{packhauser2023deep}, employ generative models to generate adversarial images, forming a new dataset for thoracic abnormality classification tasks.
Kim~\etal~\cite{kim2021privacy} proposed Privacy-Net, a framework designed to eliminate privacy-sensitive features from medical images while retaining task-relevant information for segmentation. However, these approaches often retain some visual information from the original images, which does not fully protect the visual privacy of the images. As a result, current medical analysis models face two main challenges: i) How to preserve the privacy information in medical data; and ii) How to maintain its discriminate ability after concealing the information of plaintext images within host images. 

\begin{figure}[t]
    \centering
    \includegraphics[width=1\linewidth]{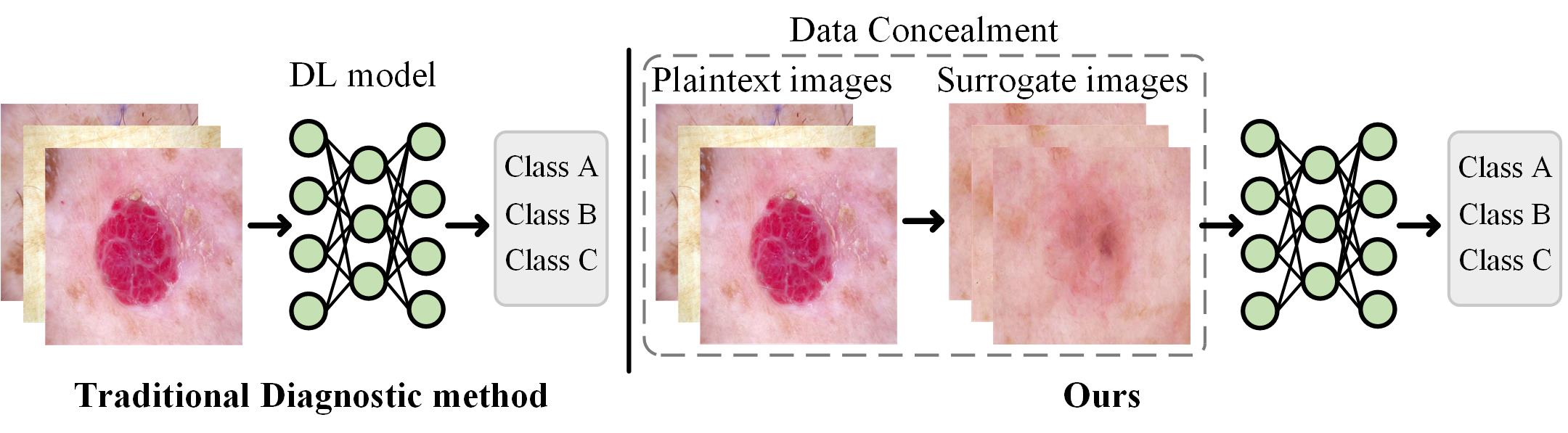}
    %\centerline{\includegraphics[width=\columnwidth]{MICCAI/fig1.jpg}}
    \vspace{-5pt}
    \caption{The concepts of the traditional medical image disease diagnosis method and ours.}
    % \vspace{-15pt}
    \label{fig1}
\end{figure}

Studies in visual psychophysics reveal that the visual recognition of images is predominantly reliant on phase information, which maintains essential visual content and perceptual quality, despite partial spectrum changes~\cite{guyader2004image,shapley1985spatial}.
Precise control of spectral transformations allows for effective information concealment while preserving the image's visual identity. 
As a result, we propose a novel Frequency-domain Exchange Style Fusion (FESF) framework to alleviate the above issues and balance the trade-off between utility and privacy.

For clarity, Fig.~\ref{fig1} illustrates the differences between the traditional medical image analysis model and ours. Traditional methods use plaintext medical images for DL model training and inference. 
In our method, we conceal plaintext medical images within host images to generate surrogate images, which are then used for training and inference in the network model. By doing this, our approach ensures no plaintext images are exposed during training and inference, greatly reducing privacy concerns related to data leakage.

%Specifically, FDEM is designed to exchange the partial frequency domains of plaintext medical images with the host images. However, the proposed concealment method would decrease the image quality, we further introduce a QEM to improve the image quality and ensure minimal visual alteration from surrogate images to the host images. 

In summary, the main contributions of this paper are highlighted as follows: 
\begin{itemize}
    \item We introduce a novel FESF framework for privacy-preserving medical image analysis by eliminating the need for plaintext images during the training and inference processes of DL models.
    %\item We preserve the data privacy by embedding plaintext medical images into host images , generating surrogate images that maintain the utility for DL-based medical image analysis models. 
    %\item We preserve the data privacy by swapping partial frequency domain information between plaintext and host medical images, then apply some image quality enhancement operations. This process ensures surrogate images retain essential plaintext data characteristics without visual alterations that maintain data utility for DL-based medical image analysis models.
    \item We preserve the data privacy by swapping partial frequency domain information between plaintext and host medical images, generating surrogate images that maintain the utility for DL-based medical image analysis models.
    \item Extensive experimental assessments across diverse datasets and with various~DL models reveal that our method effectively balances privacy and utility in medical image analysis models.
\end{itemize}

\begin{figure*}[t]
    \centering
    % \vspace{-5pt}
    \includegraphics[width=1\linewidth]{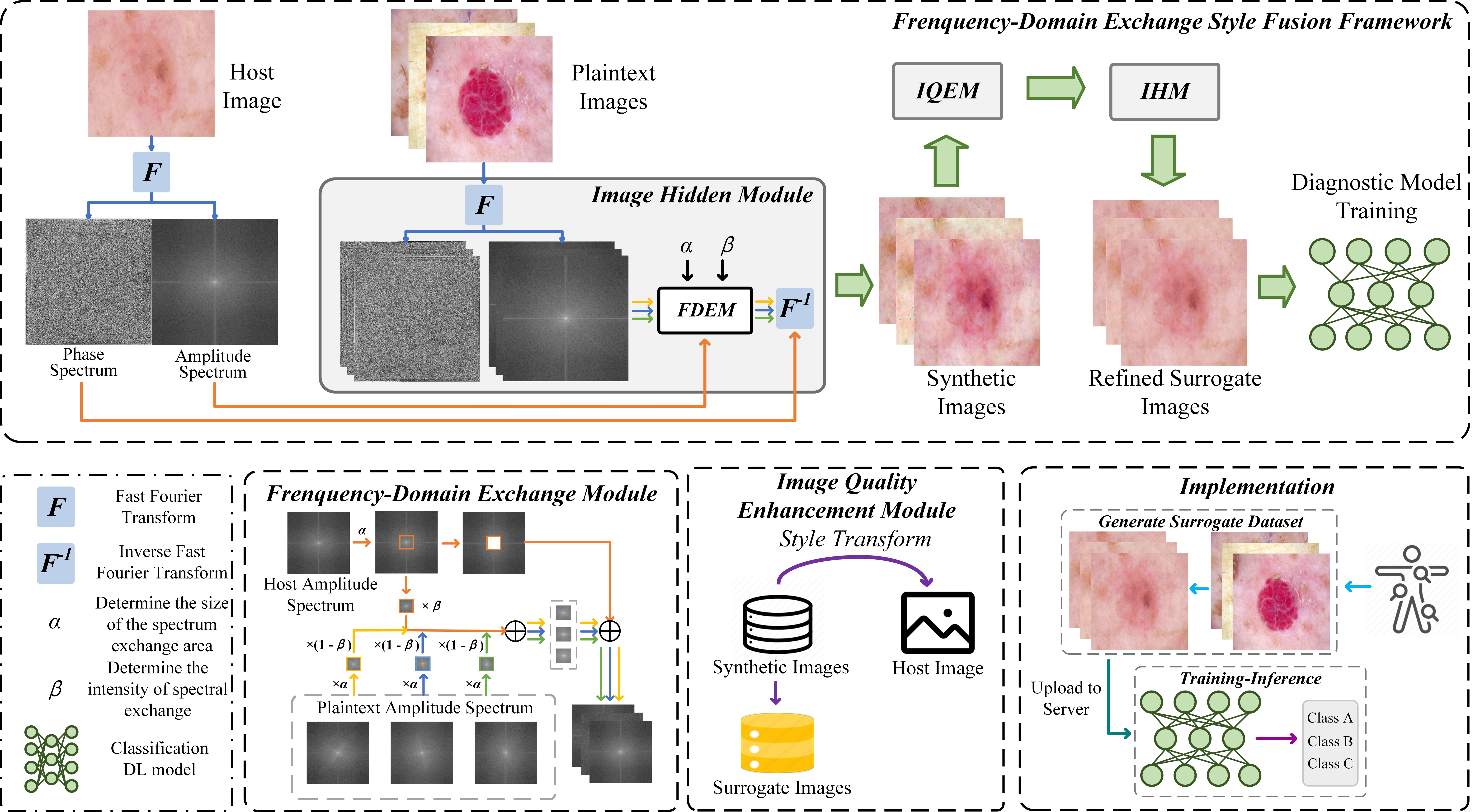}
    \caption{The overview of the proposed FESF framework.}% We use (a) Data Concealment module to conceal plaintext image into host image, and then use (b) QEM to improve the quality of surrogate images, and finally use surrogate images to train disease diagnosis DL model.}
    \label{fig2}
    \vspace{-10pt}
\end{figure*}

\section{Proposed Method}
\subsection{Overview}
The overview of the proposed FESF framework is depicted in Fig.~\ref{fig2}. The FESF framework comprises two main components: an Image Hidden Module (IHM) and an Image Quality Enhancement Module (IQEM). Initially, it transforms host images alongside plaintext images into the frequency domain using the Fast Fourier Transform (FFT), resulting in amplitude and phase spectra.
It then embeds the feature of plaintext images into the host image by manipulating specific frequency components. Subsequently, the inverse FFT generates synthetic images containing the hidden data. To further refine the visual quality of these synthetic images, the IQEM extracts features from the host images to adapt and improve the quality of the synthetic images and generates surrogate images. Then, low-intensity IHM is used to refine the surrogate images. Finally, the models are trained by feeding the surrogate images into the networks. Patients can upload their personal surrogate images to the server for diagnosis using DL models. The training and inference processes of the DL model rely solely on surrogate images, thereby eliminating the need for plaintext medical images. This framework innovatively preserves the privacy and the practical utility of sensitive medical data.

\subsection{Image Hidden Module}
The core idea of data concealment is to employ IHM to discreetly embed plaintext images within a host image. This method leverages the amplitude spectrum to represent an image's low-level distribution and the phase spectrum to encapsulate high-level semantic information~\cite{guo2023unsupervised}. Notably, changes to the amplitude spectrum have been found to minimally affect the visual perception of the image~\cite{vu2011bf}. Partial frequency domain information is extracted from plaintext images and exchanged with the counterpart of the host image by using Frequency-Domain Exchange Moddle (FDEM), ensuring that the host image conveys specific distribution details of plaintext images without noticeable visual changes. 

Initially, a host image (${x}^{ho}$) and a plaintext image dataset $X$ containing~$N$ plaintext images, which can be formulated as $X=\{x_1^{pl},...,x_N^{pl}\}$.
$N$ plaintext images %(${x_{i}^{pl}},i\in[1,N]$) 
are selected to be transformed into the frequency domain using FFT. This transformation is mathematically formulated as:

\begin{equation}
\mathcal{F}\left(x'\right)(c, m, n)=\sum_{h, w} x'(c, h, w) e^{-j 2 \pi\left(\frac{h}{H} m+\frac{w}{W} n\right)},
\label{eq:fft}
\end{equation}
where $x' \in x^{ho} \cup X \in \mathbb{R}^{C \times H\times W}$. $C$, $H$, and $W$ denotes the dimensions of the image. $(c, m, n)$ and $(c, h, w)$ represent the coordinates in frequency and spatial domains, respectively.
% $x^{k}$, $\{x_{i}^{o}\}_{i=1}^{N}\in\mathbb{R}^{H \times W \times C}$, and $C$, $H$, and $W$ index the channel ($C=3$ for an RGB image and $C=1$ for a grayscale image), height, and width of the image, respectively. $(m, n, c)$ are the indices in the frequency domain, where $m$ and $n$ index vertical (height) and horizontal (width) frequency components, respectively, and $c$ is an index variable used to iterate over individual color channels in the frequency space signal calculation. $(h, w, c)$ are the indices in the spatial domain, where $h$ and $w$ index the height and width.

This frequency space signal $\mathcal{F}\left(x'\right)$ can be further decomposed to an amplitude spectrum $\mathcal{A}_{x'}\in\mathbb{R}^{C\times H\times W}$ and a phase spectrum $\mathcal{P}_{x'}\in\mathbb{R}^{C\times H\times W}$.
% After FFT, the amplitude spectrum $\mathcal{A}$ and phase spectrum $\mathcal{P}$ of both $x_{i}^{pl}$ and ${x}^{ho}$ are obtained as:
% \begin{equation}
% \mathcal{A}_{x'}=\mathcal{F}^\mathcal{A}(x'), \mathcal{P}_{x'}=\mathcal{F}^\mathcal{P}(x'),
% \label{eq:fft1}
% \end{equation}
% where $\mathcal{F}^\mathcal{A}(\cdot)$ and $\mathcal{F}^\mathcal{P}(\cdot)$ represent the operators of amplitude and phase spectrum extractions from an image. 
%The amplitude spectrum captures the magnitude of the frequencies present in the image, while the phase spectrum preserves the positional information of these frequencies.
Then, a binary mask, $\mathcal{M}$, is used to selectively blend the low-frequency components of the host image's amplitude spectrum with those of the plaintext images. It is defined as follows:
\begin{equation}
    \mathcal{M}(m,n)=\begin{cases}
       & 1, \textit{if} \quad m \in [-\alpha H, \alpha H] \quad \textit{and} \quad n \in [-\alpha W, \alpha W] \\
       & 0, \textit{otherwise}
    \end{cases},
    \label{eq:mask}
\end{equation}
where $\alpha$ determines the low-frequency region intended for exchange. 
% The value 1 in the mask indicates the areas designated for interaction.

In order to determine the exchange ratio between the amplitude spectra of the plaintext image $\mathcal{A}_{x_i^{pl}}$ and the host image $\mathcal{A}_{x^{ho}}$, a hyper-parameter $\beta$ is introduced to modulate the intensity of the host's spectrum to construct the synthetic amplitude spectrum ${A}_{x_{i}^{sy}}$, which is defined as:
\begin{equation}
\mathcal{A}_{x_{i}^{sy}}=\left[(1-\beta) \mathcal{A}_{x_{i}^{ho}}+\beta \mathcal{A}_{x^{pl}}\right] \otimes \mathcal{M}+\mathcal{A}_{x^{ho}} \otimes (1-\mathcal{M}),
\label{eq:mask}
\end{equation}
where "$\otimes$" represents the Hadamard product operation.

Euler's formula is then employed to reconstruct the frequency domain signal using the synthetic amplitude spectrum ${A}_{x_{i}^{sy}}$ and the phase spectrum of the original host image, which is defined as $\mathcal{P}_{x^{ho}}$. Finally, the inverse FFT, denoted as $\mathcal{F}^{-1}$, is applied to yield the synthetic image $x_{i}^{sy}$. The entire synthesis process is formulated as:
% \begin
% x_{i}^{s_{0}}=\mathcal{F}^{-1}\left(\mathcal{A}_{x_{i}}^{s_{0}} * exp(1j*\mathcal{P}_{x^{k}})\right)
%   \label{eq:ifft},
% \end{equation}
% where $j$ is the imaginary unit.

\begin{equation}
x_{i}^{sy}=\mathcal{F}^{-1}\left(\mathcal{A}_{x_{i}^{sy}}, \mathcal{P}_{x^{ho}}\right).
  \label{eq:ifft}
\end{equation}

In this way, the information of the plaintext images can be effectively concealed into the host image while preserving the essential phase information of the host image. This results in a synthetic image that contains the host image's structural features.

\subsection{Image Quality Enhancement Module}
After being processed by IHM, the IQEM is introduced to enhance the visual consistency of synthetic images with the host image.
%the surrogate image exhibits a slight visual shift towards the plaintext image, which may lead to privacy leakage.
%To improve image quality and reduce the risk of visual feature leakage from plaintext images, 

% Our methodology hinges on learning bidirectional mapping functions between two distinct image domains $\mathcal{K}$ and $\mathcal{S}$, with training samples consisting of host image $x^k \in \mathcal{K}$ and surrogate images $\left\{x_{i}^{s_{0}}\right\}_{i=1}^{N} \in \mathcal{S}$. 
Our methodology aims to learn a mapping function $\mathcal{G}$ that converts images from synthetic image domain $\mathcal{S}$ to host image domain $\mathcal{K}$. An adversarial discriminator $\mathcal{D_{K}}$ is then used to distinguish between the host image $x^{ho} \in \mathcal{K}$ and the synthetic images $\{\mathcal{G}(x_{i}^{sy})\}_{i=1}^{N}, x_{i}^{sy} \in \mathcal{S}$.
% Employ an adversarial discriminator $\mathcal{D_{K}}$ to distinguish between host image $x^k$ and the transformed images $\{\mathcal{G}(x_{i}^{s_{0}})\}_{i=1}^{N}$. 
The mapping function $\mathcal{G}: \mathcal{S} \rightarrow \mathcal{K}$ and its discriminator $\mathcal{D_{K}}$ are optimized together via a min-max problem using the following adversarial loss function $\mathcal{L}_{\mathrm{GAN}}(\mathcal{G},\mathcal{D_K},\mathcal{S},\mathcal{K})$:
\begin{equation}
    \begin{aligned}
    \mathop{\min}_{\mathcal{G}}\mathop{\max}_{\mathcal{D_{K}}}\mathcal{L}_{\mathrm{GAN}}(\mathcal{G},\mathcal{D_K},\mathcal{S},\mathcal{K})=
    &\mathbb{E}_{K\sim p_{\mathrm{data}}(x^{ho})}[\log \mathcal{D_K}(x^{ho})]\\
    +&\mathbb{E}_{S\sim p_{\mathrm{data}}(x_{i}^{sy})}[\log(1-\mathcal{D_K}(\mathcal{G}(x_{i}^{sy})))],\end{aligned}\label{eq:gan}
\end{equation}
where $\mathcal{G}$ aims to generate surrogate images $\{x_{i}^{su}\}_{i=1}^{N}$ that mimic the distribution of domain $\mathcal{K}$. 
$E_{K \sim p_{\text {data }}(x^{ho})}$ and $E_{S \sim p_{\text {data }}(x_{i}^{sy})}$ denote the expectations over $\mathcal{K}$ and $\mathcal{S}$, respectively.
The primary objective of $\mathcal{G}$  is to minimize this adversarial loss, effectively fooling the discriminator into classifying generated images as belonging to domain $\mathcal{K}$. Conversely, $\mathcal{D_{K}}$ endeavors to maximize the same loss, enhancing its ability to identify the origins of the images. 
Our approach ensures that the generator $\mathcal{G}$ and discriminator $\mathcal{D_{K}}$ evolve in a symbiotic manner, promoting the generation of high-fidelity surrogate images that retain the key features of the host images.

After IQEM, we apply IHM once again to stabilize the image quality. We set $\alpha^{\prime}$ and $\beta^{\prime}$ to obtain refined surrogate images $\{x_{i}^{su'}\}_{i=1}^{N}$.

% \subsection{Diagnostic Model Training}
% In this study, four well-known networks, including VGG16~\cite{simonyan2014very}, ResNet34~\cite{he2016deep}, MobileNetv3~\cite{howard2019searching}, and EfficientNetv2~\cite{tan2021efficientnetv2} are on four surrogate datasets. 
% Each model was separately trained on the same surrogate datasets. The performance of each network was then evaluated to verify the effectiveness of our proposed framework with metrics including accuracy, precision, recall rate, and F1 score.

\subsection{Implementation}
Patients provide plaintext medical images and use IHM and IQEM to generate personal surrogate medical images. Patients then upload the surrogate images to the server, utilize a trained DL model to analyze the surrogate medical images, and return the prediction results.

 \begin{figure}[t]
      \centering
     \vspace{-5pt}
      \includegraphics[width=1\linewidth]{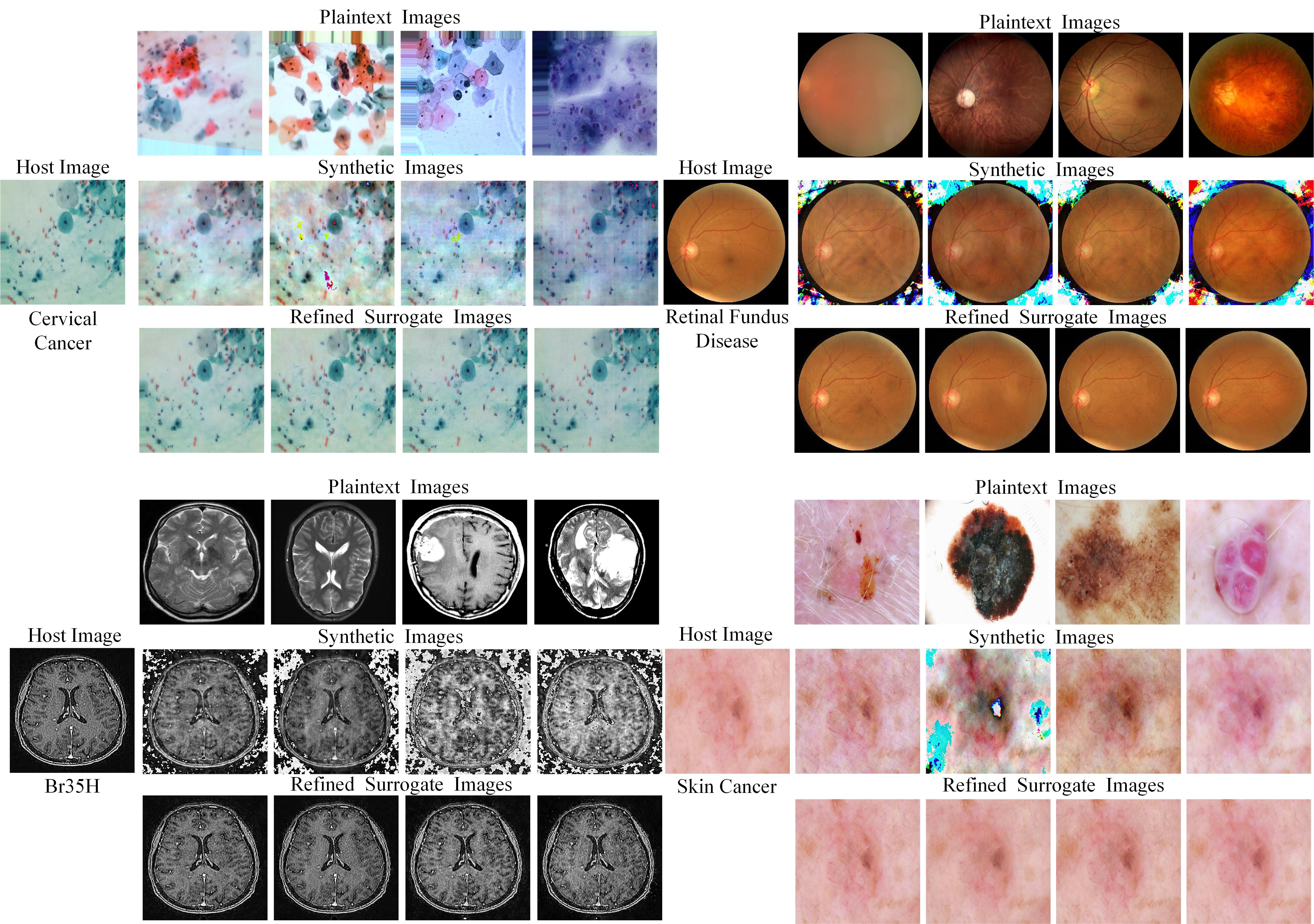}
      \caption{Visual comparison of host, plaintext, synthetic, and surrogate images across different datasets.}
      \label{fig3}
      %\vspace{-10pt}
  \end{figure}

\begin{table}[t]
% \vspace{5pt}
\caption{Quantitative analysis results of FESF framework across various medical image datasets with different DL models.}
\vspace{-8pt}
\begin{center}\resizebox{\textwidth}{!}{%
\begin{tabular}{ccccclccclccclccc}
\toprule
                                                                          &     & \multicolumn{3}{c}{\begin{tabular}[c]{@{}c@{}}Skin \\ Cancer\end{tabular}} &  & \multicolumn{3}{c}{\begin{tabular}[c]{@{}c@{}}Cervical\\ Cancer\end{tabular}} &  & \multicolumn{3}{c}{\begin{tabular}[c]{@{}c@{}}Retinal Fundus\\ Disease\end{tabular}} &  & \multicolumn{3}{c}{Br35H}         \\ \midrule
&     &  $x^{pl}$               &  $x^{su}$               & $x^{su'}$               &  &  $  x^{pl}$                & $x^{su}$                & $x^{su'}$              &  &  $x^{pl}$                  & $x^{su}$                   & $x^{su'}$                  &  & $x^{pl}$ & $x^{su}$  & $x^{su'}$ \\ \midrule
\multirow{4}{*}{\begin{tabular}[c]{@{}c@{}}Efficient-\\ Net\end{tabular}} & Acc & 83.20                   & 53.23                   & 80.36                  &  & 99.79                    & 90.59                    & 97.52                   &  & 85.38                      & 73.75                      & 74.75                      &  & 97.69     & 75.88     & 91.24     \\
                                                                          & P   & 62.06                   & 41.99                   & 59.10                  &  & 99.79                    & 90.64                    & 97.52                   &  & 77.75                      & 68.57                      & 62.00                      &  & 97.13     & 77.30     & 91.23     \\
                                                                          & R   & 63.88                   & 34.48                   & 61.62                  &  & 99.79                    & 90.59                    & 97.52                   &  & 80.47                      & 56.73                      & 72.19                      &  & 97.51     & 75.88     & 91.26     \\
                                                                          & F1  & 62.87                   & 32.85                   & 58.79                  &  & 99.79                    & 90.57                    & 97.52                   &  & 76.72                      & 59.79                      & 59.45                      &  & 96.78     & 75.76     & 91.24     \\ \midrule
\multirow{4}{*}{\begin{tabular}[c]{@{}c@{}}Mobile-\\ Net\end{tabular}}    & Acc & 85.53                   & 55.30                   & 76.74                  &  & 99.94                    &94.89                          & 98.94                   &  & 87.38                      & 79.73                      & 82.06                      &  & 98.85     & 73.37     & 92.15     \\
                                                                          & P   & 72.34                   & 43.06                   & 62.81                  &  & 99.94                    &94.89                           & 98.94                   &  & 81.10                      & 76.41                      & 70.38                      &  & 98.54     & 74.27     & 92.14     \\
                                                                          & R   & 70.70                   & 40.42                   & 64.72                  &  & 99.94                    & 94.89                          & 98.94                   &  & 82.61                      & 65.95                      & 76.53                      &  & 98.52     & 73.37     & 92.43     \\
                                                                          & F1  & 71.34                   & 40.83                   & 61.84                  &  & 99.94                    & 94.89                          & 98.94                   &  & 80.28                      & 69.35                      & 68.43                      &  & 98.55     & 73.12     & 92.15     \\ \midrule
\multirow{4}{*}{ResNet}                                                   & Acc & 88.37                   & 57.36                   & 79.59                  &  & 99.96                    & 93.72                    & 98.67                   &  & 87.71                      & 80.73                      & 83.39                      &  & 97.33     & 77.96     & 92.49     \\
                                                                          & P   & 69.73                   & 47.28                   & 62.00                  &  & 99.96                    & 93.72                    & 98.67                   &  & 81.04                      & 81.60                      & 72.15                      &  & 97.33     & 80.61     & 92.48     \\
                                                                          & R   & 71.59                   & 39.07                   & 64.33                  &  & 99.96                    & 93.72                    & 98.67                   &  & 84.00                      & 66.35                      & 80.69                      &  & 97.40     & 77.96     & 92.64     \\
                                                                          & F1  & 69.26                   & 39.41                   & 60.93                  &  & 99.96                    & 93.72                    & 98.67                   &  & 79.88                      & 69.43                      & 69.71                      &  & 97.33     & 77.48     & 92.49     \\ \midrule
\multirow{4}{*}{VGG}                                                      & Acc & 85.01                   & 59.17                   & 78.29                  &  & 99.78                    & 94.48                    & 94.72                   &  & 87.38                      &80.73                            & 81.40                      &  & 95.99     & 77.63     & 93.32     \\
                                                                          & P   & 65.16                   & 49.71                   & 58.10                  &  & 99.78                    & 94.48                    & 94.72                   &  & 78.69                      & 74.59                           & 70.46                      &  & 95.99     & 78.83     & 93.32     \\
                                                                          & R   & 65.79                   & 40.31                   & 59.56                  &  & 99.78                    & 94.48                    & 94.72                   &  & 83.44                      &66.39                            & 74.24                      &  & 96.20     & 77.63     & 93.35     \\
                                                                          & F1  & 64.98                   & 40.35                   & 58.05                  &  & 99.78                    & 94.48                    & 94.72                   &  & 78.07                      &69.41                            & 69.20                      &  & 95.99     & 77.39     & 93.32     \\   
\bottomrule
\vspace{-25pt}
\label{tabel1}

\end{tabular}}
\end{center}
\end{table}

\section{Experiments}
\subsubsection{Training Details.} 
We utilized four datasets for DL training and testing, each representing a distinct disease classification task including Br35H~\cite{Br35H}, Retinal Fundus Images~\cite{Orlando_Fu_Barbosa}, Cervical Cancer~\cite{CervicalCancer} and Skin Cancer~\cite{Skin}. Each dataset was processed using plaintext images and their corresponding surrogates generated by the FESF framework, $\alpha$ set to 0.5, $\alpha^{\prime}$ set to 0.5, $\beta$ set to 0.5, and $\beta^{\prime}$ set to 0.1.
Each dataset was randomly divided into a training set and a test set with a ratio of 6:4 and each image in the datasets was resized to $512 \times 512$ pixels. 

In this study, four well-known networks, including VGG16~\cite{simonyan2014very}, ResNet34~\cite{he2016deep}, MobileNetv3~\cite{howard2019searching}, and EfficientNetv2~\cite{tan2021efficientnetv2} are on four datasets. All the DL methods were trained with 150 epochs. The performance of each network was then evaluated to verify the effectiveness of our proposed framework with metrics including Accuracy (Acc), Precision (P), Recall rate (R), and F1 score (F1). The training was implemented using the PyTorch library and executed on one NVIDIA GeForce RTX 4090 GPU. 

% \usepackage{multirow}

\subsubsection{Experimental Results.}
As illustrated in Fig.~\ref{fig3}, plaintext images are embedded into the host images to generate surrogate images, while maintaining high similarity to the host images. Moreover, during the IQEM process, there was a notable reduction in the color feature information of the synthetic images compared to the refined surrogate images.

The quantitative results of the proposed methods across different DL networks are detailed in Tab.~\ref{tabel1}. We use the results of training only with plaintext as the upper bound. It is observed that the performance degradation is acceptable, especially in Cervical Cancer and Br35H datasets. The main reason lies in that the diagnosis of these two datasets is mainly relies on anatomical and cellular morphology features, which are preserved in the amplitude spectrum's low-frequency components. The performance degradation is greater in Retinal Fundus and Skin Cancer datasets, due to their reliance on color information for diagnosis. However, in order to preserve the privacy information and improve the image quality, IQEM would abandon kind of color information. The repeated application of IHM positively influences the utility of surrogate medical images in DL applications, potentially enhancing the models' ability to produce more accurate diagnostic outcomes.

We conduct a quantitative analysis of the host images, plaintext images, and their corresponding surrogate and refined surrogate images, with the results presented in Tab.~\ref{tabel2} and~\ref{tabel3}. 
Three image quality evaluation metrics are used to validate the image quality, including Structural Similarity Index Measure (SSIM), Peak Signal-to-Noise Ratio (PSNR), and Learned Perceptual Image Patch Similarity (LPIPS). IQEM has been shown to significantly enhance image quality, bringing the generated images into closer alignment with domain $\mathcal{K}$. 
The SSIM values for the pairs $(x^{ho},x_{i}^{su'})$ and $(x^{pl},x_{i}^{su'})$ show a marked decrease in the Cervical Cancer and Br35H datasets, suggesting substantial structural alterations in the images. 
In contrast, the Skin Cancer and Retinal Fundus Disease datasets exhibit smaller SSIM reduction due to their intrinsic similarity, which complicates classification. However, the IHM process aids in providing color information that enhances the analysis.
The PSNR of four all datasets are high which means refined surrogate images maintain a high degree of fidelity to the host images, confirming that the quality of the surrogate images is generally well-preserved. Additionally, the LPIPS scores suggest that the surrogate images hold a high degree of visual similarity to the host images, indicating effective concealment.

\begin{table}[t]
\vspace{-7pt}
\caption{Image quality analysis between host image, refined surrogate images and synthetic images.}
\centering
\begin{tabular}{ccccccc}
  \toprule
                       & \multicolumn{2}{c}{SSIM}                          & \multicolumn{2}{c}{PSNR}                          & \multicolumn{2}{c}{LPIPS}                         \\
                       &$(x^{ho},x_{i}^{su'})$ & $(x^{ho},x_{i}^{sy})$ & $(x^{ho},x_{i}^{su'})$ & $(x^{ho},x_{i}^{sy})$ & $(x^{ho},x_{i}^{su'})$ & $(x^{ho},x_{i}^{sy})$ \\ \midrule
Skin Cancer            & 0.9748                  & 0.9422                  & 34.5861                 & 28.9633                 & 0.0223                  & 0.1105                  \\
Cervical Cancer        & 0.9297                  & 0.8679                  & 31.5723                 & 28.4355                 & 0.0488                  & 0.1934                  \\
Retinal Fundus Disease & 0.9005                  & 0.6851                  & 32.6282                 & 28.8184                 & 0.0638                  & 0.3170                  \\
Br35H                  & 0.8280                  & 0.4477                  & 30.3095                 & 28.1640                 & 0.1094                  & 0.5375                 \\
\bottomrule
%\vspace{-15pt}
\label{tabel2}
\end{tabular}
\end{table}

\begin{table}[t]
\vspace{-10pt}
\caption{Image quality analysis between plaintext images, refined surrogate images and synthetic images.}
\centering
\begin{tabular}{ccccccc}
  \toprule
                       & \multicolumn{2}{c}{SSIM}                          & \multicolumn{2}{c}{PSNR}                          & \multicolumn{2}{c}{LPIPS}                         \\
                       &$(x^{pl},x_{i}^{su'})$ & $(x^{pl},x_{i}^{sy})$ & $(x^{pl},x_{i}^{su'})$ & $(x^{pl},x_{i}^{sy})$ & $(x^{pl},x_{i}^{su'})$ & $(x^{pl},x_{i}^{sy})$ \\ \midrule
Skin Cancer            & 0.7066                  & 0.6801                  & 28.1574                & 28.4204                & 0.3807                 & 0.3126                  \\
Cervical Cancer        & 0.4976                  & 0.4664                  & 27.9686                & 28.0364                 & 0.5485             & 0.5314              \\
Retinal Fundus Disease & 0.6488                 & 0.4975                  & 29.0442                 & 28.4616                 & 0.3807             & 0.4822                 \\
Br35H                  & 0.1352                 & 0.0815                 & 28.3169                & 28.2764                & 0.5195                  & 0.6510                 \\
\bottomrule
\vspace{-15pt}
\label{tabel3}
\end{tabular}
\end{table}

% \subsubsection{Imperceptivity Analysis.} "Imperceptibility" refers to the concealment capability of an information hiding system to prevent perceptible degradation in the host image during the embedding process. 
% As illustrated in Fig.~\ref{fig3}, plaintext images are embedded into the host images, generating surrogate images while maintaining high similarity to the host images.
% To quantitatively measure the imperceptivity of surrogate images, we apply Peak Signal-to-Noise Ratio (PSNR), Structural Similarity Index Measure (SSIM), and Learned Perceptual Image Patch Similarity (LPIPS) as metrics. 

% A strong correlation exists between the similarity metrics of surrogate and host images and the effectiveness of information hiding. 
% Tabs.~\ref{tabel2} and~\ref{tabel3} show that surrogate images have higher SSIM values compared to host images, suggesting a favorable structural match. 
% High PSNR values indicate minimal distortion, while low LPIPS values confirm perceptual similarity. 
% Collectively, these metrics confirm the effectiveness of our method in hiding plaintext images within host images. Moreover, the utilization of IQEM substantially enhances the concealment of plaintext images within surrogate images.

\section{Conclusion}
In this study, we propose a novel FESF framework specifically designed to address critical data privacy concerns in the application of DL in the medical field, particularly during model training and inference stages. We ensure data privacy by using the IHM to extract partial frequency domain information from plaintext medical images and exchange this information with the counterpart of the host image. 
This process guarantees that the surrogate images convey certain distribution details from the plaintext images while visually unrelated to the plaintext image.  Extensive experiments on various datasets with different DL models demonstrate that our framework offers two main benefits: surrogate images can replace plaintext images and maintain diagnostic accuracy of DL models at a relatively high level, and it well balances privacy and utility. %In future work, 
\bibliographystyle{splncs04}
\bibliography{HIDDEN_IMAGE}
\end{document}